\documentclass[prb,twocolumn,showpacs,amsmath,amssymb]{revtex4-1}
\usepackage{graphicx}
\DeclareGraphicsExtensions{.png,.jpg,.eps}
\usepackage{dcolumn}
\usepackage{bm}
\usepackage[latin1]{inputenc}
\usepackage{xcolor}
\usepackage{float}
\usepackage{amsmath}

\newcommand{\sect}[1]{\emph{#1.---}\ignorespaces}

\begin{document}
\date{\today}

\title{Electrically controllable magnetism in twisted bilayer graphene}
\author{Luis A. Gonzalez-Arraga$^1$, J. L. Lado$^2$, Francisco Guinea$^{1,3}$ and Pablo San-Jose$^4$}
\affiliation{$^1$IMDEA Nanociencia, Calle de Faraday, 9, Cantoblanco, 28049, Madrid, Spain}
\affiliation{$^2$International Iberian Nanotechnology Laboratory (INL), Av. Mestre Jose Veiga, 4715-330 Braga, Portugal}
\affiliation{$^3$School of Physics and Astronomy, University of Manchester, Oxford Road, Manchester M13 9PL, UK}
\affiliation{$^4$Instituto de Ciencia de Materiales de Madrid (ICMM-CSIC). Cantoblanco, 28049 Madrid, Spain}
 
\begin{abstract}
Twisted graphene bilayers develop highly localised states around AA-stacked
regions for small twist angles. We show that interaction effects may induce
either an antiferromagnetic (AF) and a ferromagnetic (F) polarization of said
regions, depending on the electrical bias between layers. Remarkably,
F-polarised AA regions under bias develop spiral magnetic ordering, with a
relative $120^\circ$ misalignment between neighbouring regions due to a
frustrated antiferromagnetic exchange. This remarkable spiral magnetism emerges
naturally without the need of spin-orbit coupling, and competes with the more
conventional lattice-antiferromagnetic instability, which interestingly
develops at smaller bias under weaker interactions than in monolayer graphene,
due to Fermi velocity suppression. This rich and electrically controllable
magnetism could turn twisted bilayer graphene into an ideal system to study frustrated magnetism in two dimensions, with interesting potential also for a range of applications.
\end{abstract}

\maketitle


{
Magnetism in two dimensional (2D) electronic systems is known to present
very different phenomenology from its three-dimensional counterpart due to the reduced dimensionality and the increased importance of
fluctuations. Striking examples are the impossibility of establishing long range magnetic order in a 2D system without magnetic anisotropy \cite{Mermin:PRL66} or the emergence of unique finite-temperature phase transitions that are controlled by the proliferation of topological magnetic defects \cite{Nelson:PRL77}. In the presence of magnetic frustration, in e.g. Kagome \cite{Fu:S15,Lee:NM07} or triangular lattices \cite{Isono:PRL14,Seabra:PRB11,Hu:PRB15a,Zhu:PRB15}, 2D magnetism may also lead to the formation of remarkable quantum spin liquid phases \cite{Fu:S15,Savary:ROPIP16,Xu:PRL16}. The properties of these states remain under active investigation, and have recently been shown to develop exotic properties, such as fractionalized excitations \cite{Han:N12}, long-range quantum entanglement of their ground state \cite{Grover:NJOP13,Pretko:PRB16}, topologically protected transport channels \cite{Yao:NC13} or even high-$T_C$ superconductivity upon doping\cite{Lee:NM07,Kelly:PRX16,Anderson:MRB73}.} 

{The importance of 2D magnetism extends also beyond fundamental physics into applied fields. One notable example are data storage technologies. Recent advances in this field are putting great pressure on the magnetic memory industry to develop solutions that may remain competitive in speed and data densities against new emerging platforms. Magnetic 2D materials are thus in demand as a possible way forward \cite{Wang:2M16}. Of particular interest for applications in general are 2D crystals and van-der-Waals heterostructures. These materials have already demonstrated great potential for a wide variety of applications, most notably nanoelectronics and optoelectronics \cite{Novoselov:S16,Castellanos-Gomez:NP16,Quereda:NL16}. Some of them have been shown to exhibit considerable tuneability through doping, gating, stacking and strain. Unfortunately, very few 2D crystals have been found to exhibit intrinsic magnetism, let alone magnetic frustration and potential spin liquid phases.}

{In this work we predict that twisted graphene bilayers could be a notable exception, realizing a peculiar magnetism on an effective triangular superlattice, and with exchange interactions that may be tuned by an external electric bias}. We show that spontaneous magnetization of two different types may develop for small enough twist angles $\theta\lesssim 2^\circ$ as a consequence of the moir\'e pattern in the system. This effect is a consequence of the high local density of states generated
close to neutrality at moir\'e regions with AA stacking, triggering a Stoner
instability when electrons interact.  The local order is localized at AA regions but may be either antiferromagnetic (AF) or ferromagnetic (F). The two magnetic orders can be switched electrically by applying a voltage bias between layers. Interestingly
the relative ordering between different AA regions in the F ground state is predicted
to be \emph{spiral}, despite the system possessing negligible spin-orbit
coupling. {The magnetism of the system thus combines a set of unique features: electric tuneability, magnetic frustration, interplay of two switchable magnetic phases with zero net magnetization, spatial localization of magnetic moments, and an adjustable period of the magnetic superlattice. These make twisted graphene bilayers a prime playground for studies into spin liquid phases, and for potential applications such as magnetic memories. We discuss some of these possibilities in our concluding remarks.}

\begin{figure}
   \centering
   \includegraphics[width=\columnwidth]{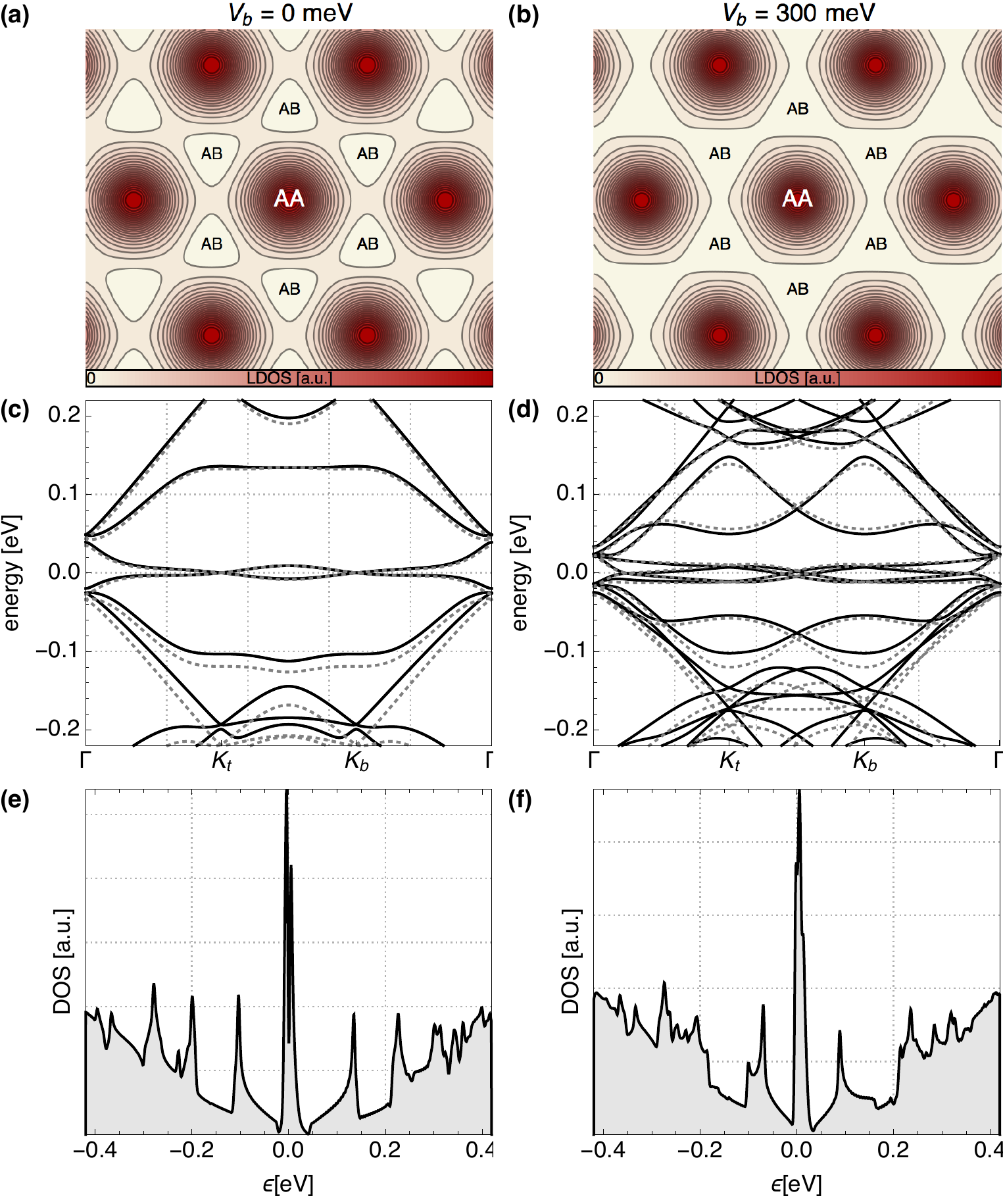}
   \caption{Zero-energy local density of states in real space (a,b), bandstructure (c,d) and density of states (e,f) for a $\theta=1.5^\circ$ twisted graphene bilayer. The left column has no interlayer bias, and the right column has a bias $V_b=300$ meV. This enhances the localization of the AA quasibound states, red in (a,b). Said states arise from almost flat subbands at zero energy, which show up as large DOS peaks in (e,f). Solid (dashed) lines in (c,d) correspond to a scaled (unscaled) tight-binding model, see main text.}
   \label{fig:elstruct}
\end{figure}

\sect{Description of the system} Twisted graphene bilayers are characterized by
a relative rotation angle $\theta$ between the two layers \cite{Santos:PRL07}.
The rotation produces a modulation of the relative stacking at each point,
following a moir\'e pattern of period $L_M\approx a_0/\theta$ at small
$\theta$, where $a_0=0.24$ nm is graphene's lattice constant
\cite{Santos:PRB12}. The stacking smoothly interpolates between three basic
types, AA (perfect local alignment of the two lattices) and AB/BA (Bernal
stackings related by point inversion) \cite{Alden:PNAS13}. The stacking
modulation leads to a spatially varying coupling between layers. This results
in a remarkable electronic reconstruction at small angles $\theta\lesssim
1-2^\circ$ \cite{Laissardiere:NL10,Bistritzer:PNAS11}, for which the interlayer
coupling $\gamma_1\sim 0.3$eV exceeds the moir\'e energy scale $\epsilon_M =
\hbar v_F \Delta K$ (here $\Delta K=4\pi/[3L_M]$ is the rotation-induced
wavevector shift between the Dirac points in the two layers, and $v_F\approx
10^6 m/s$ is the monolayer Fermi velocity). It was shown
\cite{Santos:PRB12,Bistritzer:PNAS11, San-Jose:PRL12, Laissardiere:PRB12,
Sboychakov:PRB15} that in such regime, the Fermi velocity of the bilayer
becomes strongly suppressed, and the local density of states close to
neutrality becomes dominated by quasilocalized states in the AA regions
\cite{Laissardiere:NL10}. The confinement of these states is further enhanced
by an interlayer bias $V_b$, which effectively depletes the AB and BA regions
due to the opening of a local gap. At sufficiently small angles this was also
shown to result in the formation of a network of helical valley currents
flowing along the boundaries of depleted AB and BA regions
\cite{San-Jose:PRB13}.

The quasilocalised AA-states form a weakly coupled triangular superlattice of period $L_M$, analogous to a network of quantum dots. Each AA `dot' has space for eight degenerate electrons, due to the sublattice, layer and spin degrees of freedom. A plot of their spatial distribution under zero and large bias $V_b=300$ meV is shown in Figs. \ref{fig:elstruct}(a,b), respectively. These AA states form a quasi-flat band at zero energy \cite{Suarez-Morell:PRB10}, see panels (c,d), which gives rise to a zero-energy peak in the density of states (DOS). The small but finite width of this zero-energy AA resonance represents the residual coupling between adjacent AA dots due to their finite overlap. A comparison of panels (a,b) shows that a finite interlayer bias leads to a suppression of said overlap and a depletion of the intervening AB and BA regions, as described above.
The electronic structure presented here was computed using the tight-binding approach described in the Appendix, which includes a scaling approximation that allows the accurate and efficient computation of the low-energy bandstructure in low-angle twisted bilayers (compare solid and dashed curves in panels [c,d]). Our scaling approach makes the problem much more tractable computationally, which is a considerable advantage when dealing with the interaction effects, discussed below.

\sect{Moir\'e-induced magnetism} It is known that in the presence of
sufficiently strong electronic interactions, a honeycomb tight-binding lattice
may develop a variety of ground states with spontaneously broken symmetry
\cite{Sorella:EEL92, Herbut:PRL06, Sorella:SR12, Assaad:PRX13,
Garcia-Martinez:PRB13}. The simplest one is the lattice antiferromagnetic phase
in the honeycomb Hubbard model. The Hubbard model is a simple description
relevant to monolayer graphene with strongly screened interactions (the
screening may arise intrinsically at high doping or e.g. due to a metallic
environment). Above a critical value of the Hubbard coupling,
$U>U_c^{(0)}\approx 5.7eV$ (value within mean field), the system favours a
ground state in which the two sublattices are spin-polarized
antiferromagnetically. This is known as lattice-AF (or N\'eel) order.

In the absence of adsorbates \cite{Gonzalez-Herrero:S16}, edges
\cite{Magda:N14}, vacancies \cite{Palacios:PRB08} or magnetic flux
\cite{Young:NP12} isolated graphene monolayers, with their vanishing density of states at low energies, are known experimentally not to
suffer any interaction-induced magnetic instability.  
{In contrast, Bernal ($\theta=0$) bilayer graphene and ABC trilayer graphene
have been suggested \cite{Bao:NP11,Lee:NC14,Velasco:NN12,Kharitonov:PRB12b} to develop
magnetic order, due to their finite low-energy density of states, although some controversy remains \cite{Castro:PRL08,
Nandkishore:09,Vafek:PRB10,Mayorov:S11,Lemonik:PRB12,Throckmorton:PRB14}.}
{Twisted
graphene bilayers at small angles exhibit an even stronger enhancement of the low-energy density of states associated to AA-confinement and the formation of quasi-flat bands. It is thus natural} to expect some form
of interaction-induced instability in this system with
realistic interactions, despite the lack of magnetism in the
monolayer. By analysing the Hubbard model in twisted bilayers we now explore this
possibility, and describe the different magnetic orders that emerge in the $U,
V_b$ parameter space. 

We consider the Hubbard model in a low angle $\theta\approx 1.5^\circ$ twisted
bilayer for a moderate value of $U=3.7$, quite below the monolayer lattice-AF
critical interaction $U_c^{(0)}$. We use a self-consistent mean-field
approximation to compute the system's ground state, and use the same parameters
of Fig. \ref{fig:elstruct}. Self-consistency involves the iterative computation
of charge and spin density on the moir\'e supercell, integrated over Bloch
momenta, see the Appendix for details. In Fig.
\ref{fig:mag} we show the resulting real-space distribution of the ground-state
spin polarization $M(\vec r)$ of the converged solution. The top and bottom
rows correspond, respectively, to the lattice-F and lattice-AF components
$M_A+M_B$ and $M_A-M_B$, where the polarization density is defined as
$M_{\tau}=\sum_\lambda\langle n_{\uparrow \tau\lambda}(\vec r)-n_{\downarrow
\tau\lambda}(\vec r)\rangle$. Here $\tau=A,B$ are the two sublattices and
$\lambda=\pm$ are the two layers.

\begin{figure}
   \centering
   \includegraphics[width=\columnwidth]{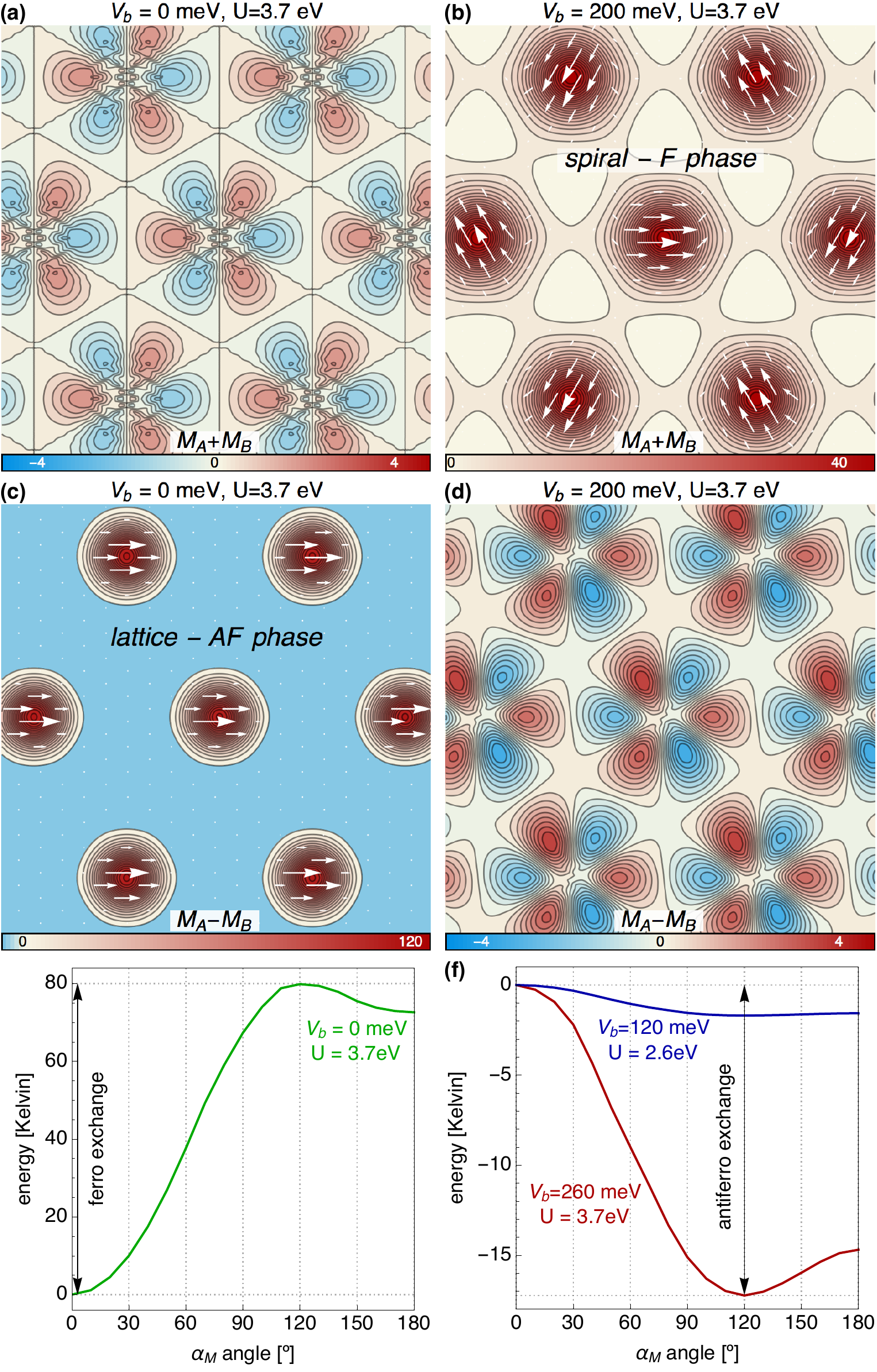}
   \caption{Spatial distribution of the magnetic moment in the ground state of
an interacting twisted bileyer with Hubbard $U=3.7$eV. In the first row (a,b)
we show the ferromagnetic component of the two sublattices, $M_A+M_B$ in units
of electrons per (monolayer) unit cell, both for zero interlayer bias $V_b=0$
(a) and $V_b=200$ meV (b). Analogous plots of the lattice-AF component
$M_A-M_B$ are shown in (c,d). The scale in all color bars is expressed in units
of one electron spins per supercell. Panels (e,f) show the variation of total
electronic energy per supercell as a function of the angle $\alpha_M$ between
polarizations of adjacent AA regions, indicating parallel alignment of the
lattice-AF order (e), and a spiral misalignment of $120^\circ$ for the
lattice-F case (f). }
   \label{fig:mag} 
\end{figure}

\begin{figure}
   \centering
   \includegraphics[width=\columnwidth]{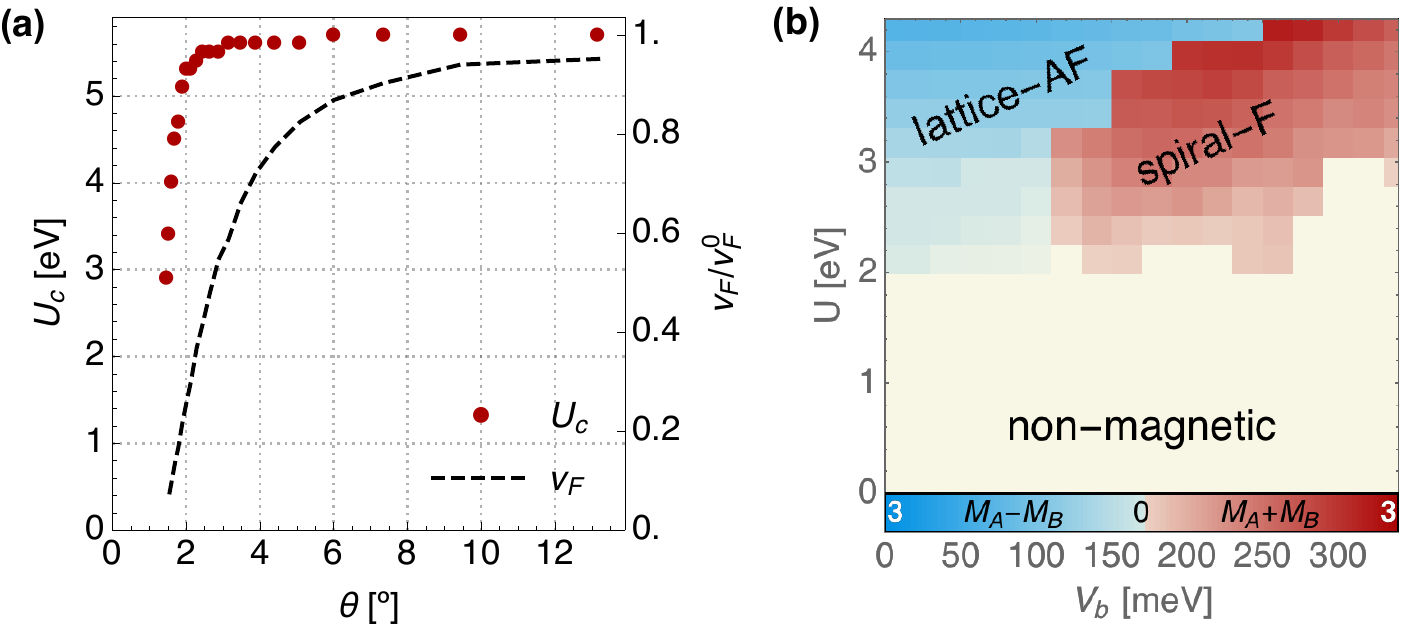}
   \caption{(a) Critical value $U_c$ of the Hubbard U beyond which the twisted
bilayer develops lattice-AF order {at the mean field level}.
Red dots show $U_c$ as a function of twist
angle $\theta$, and the dashed line show the corresponding Fermi velocity at
the Dirac point, normalized to the monolayer value $v_F^0$. At high twist
angles both $U_c$ and $v_F$ converge to the monolayer values, while they become
strongly suppressed at smaller angles. (b) Phase diagram for the ground state
magnetic order in a $\theta=1.5^\circ$ twisted bilayer as a function of Hubbard
U and interlayer bias $V_b$. Blue and red regions denote the spatial integral
of the lattice-AF and spiral-F polarizations, respectively, while the yellow
region is non-magnetic.}
   \label{fig:Ucphase} 
\end{figure}

We obtain two distinct solutions for the magnetization, depending on the
interlayer bias $V_b$. At small interlayer bias and for the chosen $U=3.7$ 
{eV} we
see that the  ferromagnetic polarization (panel [a] in Fig. \ref{fig:mag}) is
small and collinear, and spatially integrates to zero. Thus, the unbiased
bilayer remains non-ferromagnetic in the small $V_b$ case. However, the
lattice-AF component of the polarization, panel (c), is large and integrates to
a non-zero value of around $0.5$ electron spins per unit cell. This is the
analogue of the monolayer lattice-AF phase, with two important differences. On
the one hand, we find that the lattice-AF density is strongly concentrated at
the AA regions instead of being spatially uniform like in the monolayer. On the
other hand the lattice-AF ground state is found to arise already for $U\gtrsim
2$eV, i.e. for much weaker interactions than in the monolayer. The reason for
the reduction of $U_c$ can be traced to the suppression of the Fermi velocity
$v_F$ at small twist angles \cite{Bistritzer:PNAS11, Laissardiere:PRB12}, which
controls the critical $U$ for the lattice-AF instability. The dependence of
$U_c$ and $v_F$ as a function of angle $\theta$ is shown in Fig.
\ref{fig:Ucphase}(a). This result already points to strong magnetic
instabilities of twisted graphene bilayers as the angle falls below the
$1-2^\circ$ threshold.

Under a large electric bias between layers, the ground state magnetization for
the same $U$ is dramatically different, see panels (b,d) of Fig. \ref{fig:mag}.
In this case, the lattice-AF polarization, panel (d), is strongly suppressed
and integrates to zero spatially, while the lattice-F component, panel (b),
becomes large around the AA regions, and integrates to a finite value of
approximately 4 electron spins per moir\'e supercell. The AA regions are thus
found to become ferromagnetic under sufficient interlayer bias. This type of
magnetic order is the result of the increased confinement of AA states at high
$V_b$, and can be interpreted as an instance of flat-band ferromagnetism driven
by the Stoner mechanism. 

The lattice-AF and lattice-F states are also different when comparing the
relative orientations of neighbouring AA regions. By computing the total energy
per supercell in each case as a function of the polarization angle $\alpha_M$
between adjacent regions (panels [e,f] of Fig. \ref{fig:mag}), we find that the
energy is minimized for $\alpha_M=0^\circ$ in the lattice-AF case (parallel
alignment), but for  $\alpha_M=120^\circ$ in the lattice-F case (spiralling
polarization). The equilibrium polarization is depicted by white arrows in
Figs. \ref{fig:mag}(c,b). The depth of the energy minimum, ranging from $\sim
2-100$ Kelvin in our simulations, represents the effective exchange coupling of
neighboring AA regions, which is ferromagnetic for lattice-AF states and
antiferromagnetic for lattice-F states. In the latter, which from now on we
denote spiral-F phase, the spiral order arises as a result of the triangular
symmetry of AA regions that frustrates a globally antiferromagnetic
AA-alignment.
The same spiral order has been described in studies of the
Hubbard model in the triangular lattice. It is a rather remarkable magnetic
state, as the polarization at different points becomes 
non-collinear
\cite{Hu:PRB15a,Bernu:PRL92,Capriotti:PRL99} despite
the complete absence of spin-orbit coupling in the system. It should be noted that global spiral order is strictly a ground state (zero temperature) property. At finite temperature, spin excitations (gapless Goldstone modes in the magnetically isotropic case under study) are expected to destroy long-range spiral order, which then survives only locally, in keeping with the Mermin-Wagner theorem \cite{Mermin:PRL66}.


To better understand the onset of the spiral magnetism, we have computed the
integrated F and AF polarization across the $U,V_b$ plane. We find sharp
first-order phase transitions separating the two types of ground states. The
result is shown in Fig. \ref{fig:Ucphase}. Regions in red and blue denote,
respectively, a finite spatial integral of the ferro $M_A+M_B$ and lattice-AF
$M_A-M_B$ polarizations. It can be seen that an electric interlayer bias of
around 120 meV is able to switch between the lattice-AF and spiral F orders for
values of $U$ between 2 and 3eV. The precise thresholds for such electric
switching of magnetic order depend on the specific twist angle and of further
details not considered in this work (e.g. longer-range interactions,
spontaneous deformations or interlayer screening), but our simulations suggest
that it is likely to be within reach of current experiments for sufficiently
small $\theta$.

\sect{Conclusion} For a long time unmodified graphene was thought to be
relatively uninteresting from the point of view of magnetism. Twisted graphene
bilayers, however, could prove to be a surprisingly rich playground for
non-trivial magnetic phases. We have shown that two different types of magnetic
order arise spontaneously in twisted graphene bilayers at small angles. We
identified two types of magnetic order, lattice-antiferromagnetism and
spiral-ferromagnetism, both concentrated at AA-stacked regions. The spiral-F
phase is favoured over the lattice-AF when applying a sufficient electric bias
between layers. This phase constitutes a form of electrically-controllable, non-collinear and spatially non-uniform magnetism in a material with a negligible spin-orbit coupling.

{This possibility is of fundamental interest, as it realises electrically tuneable 2D magnetism on a triangular superlattice, a suitable platform to explore spin-liquid phases. Indeed, it is known that next-nearest neighbour interactions in magnetic triangular lattice should transform spiral order into a spin-liquid phase \cite{Isono:PRL14,Seabra:PRB11,Hu:PRB15a,Zhu:PRB15}, as long as the system remains magnetically isotropic. If the magnetic isotropy is broken, e.g. though a magnetic substrate which could favor parallel and antiparallel orientations of the lattice-AF phase through sublattice polarization, long-range magnetic order could be stabilized. This system could then become useful for magnetic storage applications, with one bit per antiferromagnetic AA region. In this regard it exhibits a number of desireable features, such as very high data density (given by the moir\'e period), potential immunity to neighboring bit flips (due to the zero stray fields of the lattice-AF order \cite{Loth:S12}), electrically controllable write processes (e.g. by switching a given AA region to be written from antiferro to ferro, followed by a magnetic pulse), and even purely electrical readout (due to the topologically protected spin-valley currents that arise along the boundary of opposite AF regions). While the above is highly speculative at this point and would require a detailed analysis, it highlights the interesting fundamental and practical possibilities afforded by the rich magnetic phase diagram of twisted graphene bilayers.} 


We acknowledge financial support from the Marie-Curie-ITN programme through Grant No. 607904-SPINOGRAPH, and the Spanish Ministry of Economy and Competitiveness through Grant No. FIS2015-65706-P (MINECO/FEDER) and RYC-2013-14645 (Ram\'on y Cajal programme). L. G.-A. thanks the hospitality of the Applied Physics Department in the University of Alicante and N. Garcia for useful discussions. We specially thank J. Fernandez Rossier for his help settling the environment and the initial idea for this work.

\appendix
\section{Tight-binding model for twisted graphene bilayers. Re-escaling}

The twisted bilayer graphene (TBG) lattice consists of two super-imposed graphene lattices rotated by an angle $\theta$ separated by a distance $d=3.35$ A . We label the bottom (top) monolayer by 1 (2). The carbon atoms of the monolayer 1 are located in positions given by the vectors: 

\begin{equation}
\vec{r}_{n,m}^{1A}=n \vec{a_1}+m \vec{a_2}
\end{equation}
\begin{equation}
\vec{r}_{n,m}^{1B}=n \vec{a_1}+m \vec{a_2}+\vec{\delta_1}
\end{equation}
$n$ and $m$ are integers, $\vec{\delta_1}$ is the vector separating the  $A$ and $B$ sublattices and $\vec{a_1}$  and $\vec{a_2}$ are the lattice vectors of graphene: 
\begin{equation}
\vec{a}_1= a\left(\frac{\sqrt{3}}{2}\hat{x}-\frac{1}{2}\hat{y}\right)
\end{equation}
\begin{equation}
\vec{a}_2= a\left(\frac{\sqrt{3}}{2}\hat{x}+\frac{1}{2}\hat{y}\right)
\end{equation}
\begin{equation}
\vec{\delta}_1=\frac{\vec{a_1}+\vec{a_2}}{3}
\end{equation}
The positions of atoms in monolayer 2 are given by: 
\begin{equation}
\vec{r}_{n,m}^{2B}=n \vec{a_1}^{\prime}+m \vec{a_2}^{\prime}
\end{equation}
\begin{equation}
\vec{r}_{n,m}^{2A}=\vec{r_{n,m}}^{2B}-\vec{\delta_2}
\end{equation}
where $\vec{a}_1^{\prime}$ and $\vec{a_2}^{\prime}$  are given by:
\begin{equation}
\vec{a}_1^{\prime}=\left( \cos\theta-\frac{\sin\theta}{\sqrt{3}}\right)\vec{a_1}+\frac{2\sin\theta}{\sqrt{3}}\vec{a_2}
\end{equation}
\begin{equation}
\vec{a}_2^{\prime}=\left( \cos\theta+\frac{\sin\theta}{\sqrt{3}}\right)\vec{a_2}-\frac{2\sin\theta}{\sqrt{3}}\vec{a_1}
\end{equation}
and $\vec{\delta_2}=\left( \frac{\vec{a_1}^{\prime}+\vec{a_2}^{\prime}}{3}\right)$. For an arbitrary value of $\theta$ the structure is generally incommensurate, and no unit cell can be constructed. The twisted bilayer graphene forms periodic Moire patterns only for specific $\theta$ angles that satisfy the condition:
\begin{equation}
\cos\theta=\frac{3 m_{0}^2+3m_{0}r+r^2/2}{3 m_{0}^2+3m_{0}r+r^2}
\end{equation}  
with $m_0$ and $r$ are coprime positive integers \cite{Santos:PRL07,Santos:PRB12,Sboychakov:PRB15}. The number of atoms in the Moire unit cell is given by $N(m_0,r)=4(3{m_0}^2+3m_0 r +r^2)$. The lattice vectors of the superlattice are: 
\begin{equation}
\vec{R}_1=m_0 \vec{a}_1+ (m_0+r)\vec{a}_2
\end{equation}
\begin{equation}
\vec{R}_2=-(m_0+r) \vec{a}_1+ (2m_0+r)\vec{a}_2
\end{equation}

We consider a tight-binding Hamiltonian for the $p_z$ orbitals of the carbon atoms in the lattice: 
\begin{equation}
H=\sum_{i,j} -t_{i,j}(\vec{r_{ij}}){c_{j}}^{\dagger}c_{i}+\sum_{i} V_{i} c^{\dagger}_{i} c_{i} 
\end{equation}
where $c_{i}$ destroys an electron in the $p_z$ orbital of the $i$-th site and ${c_{j}}^{\dagger}$ creates an electron in the $p_z$ orbital of the $j$-th site, $\vec{r}_{ij}=\vec{r}_i -\vec{r}_j=(x,y,z)$ is the vector separating the $i$-th and $j$-th site. The interlayer bias $V_{i}$ is an onsite energy term with opposite sign in monolayers 1 and 2. The hopping parameter $t_{i,j}(\vec{r}_{ij})$  takes into account the fact that the distances between the atoms of the different monolayers are all different. The hopping function \cite{Suarez-Morell:PRB10,Laissardiere:NL10,Moon:PRB13,Moon:PRB14} is:
\begin{eqnarray}
t_{i,j}(\vec{r_{ij}})=\gamma_{0}\exp\left[ \beta\left( \frac{r_{ij}}{a_{cc}}-1\right)\right] \left( \frac{x^2+y^2}{r_{ij}^2}\right) \\ \nonumber
+\gamma_{1}\exp\left[ -\beta\left( \frac{r_{ij}-d}{a_{cc}}\right)\right] \frac{z^2}{r_{ij}^2}
\end{eqnarray}
where $\gamma_0=-2.70$ eV is the hopping between nearest-neighbors in the same monolayer and $\gamma_1 = 0.48$ eV is the hopping between atoms belonging to different monolayer that are on top of each other. $\beta=3.137$ is a dimensionless exponential decay factor. Hoppings between atoms for $r_{ij} > 4 a_{cc}$ are negligible. 

The Brillouin zone of the monolayers of the Moire superlattice are also rotated by an angle $\theta$ and their respective K points are separated by a distance $\Delta K(\theta)=\frac{4\pi}{3\sqrt{3}a_{cc}} 2 \sin{\theta/2}$ in momentum space. The Dirac cones of the monolayers intersect in the M point of the Brillouin zone of the twisted bilayer superlattice. This intersection is observed as low-energy van-Hove singularites in the total density of states of the superlattice. In the low $\theta$ limit, $\Delta K$ becomes increasingly small and the Dirac cones intersect around an energy smaller than $\gamma_1$. This happens for $\theta \lesssim 1-2 ^{\circ} $, so that for smaller angles a flat band is formed around the Dirac point. We concentrate on an angle $\theta=1.47^\circ$ close to this threshold, corresponding to $r=1$, $m_0=22$. Our main goal is to study the magnetic order in the mean-field limit originating from the electron confinement in the AA-stacking. This iterative self-consistent approach is extremely time-consuming since the unit cells for these angles contain more than 5000 atoms. Our strategy is therefore to perform a re-escaling, in which low-energy electronic structure of the small-angle limit can be reproduced with a unit cell containing a smaller number of atoms (and larger twisting angle $\theta'$), while keeping invariant the two most important observables: the Fermi velocity and Moire period. This can be accomplished by the following scaling transformation: 
\begin{equation}
\gamma_{0}^{\prime} \rightarrow \frac{1}{\lambda}\gamma_{0}
\end{equation}
\begin{equation}
a_{cc}^{\prime} \rightarrow \lambda a_{cc},
\end{equation}
\begin{equation}
d^{\prime} \rightarrow \lambda d,
\end{equation}

where the dimensionless re-escaling parameter $\lambda$ is given by:
\begin{equation}
\lambda= \frac{\sin{\frac{\theta^{\prime}}{2}}}{\sin{\frac{\theta}{2}}}
\end{equation}

\section{Mean-field solutions}

In this section, we give a detailed explanation of the electron-electron repulsion terms included in our model. The tight-binding Hamiltonian now includes an interaction term,
\begin{equation}
H=\sum_{i,j} -t_{i,j}(\vec{r_{ij}}){c_{j}}^{\dagger}c_{i}+\sum_{i} V_{i} c^{\dagger}_{i} c_{i} +U\sum_{i} n_{i\uparrow} n_{i\downarrow}
\end{equation}

To compute the expected electronic structure for finite $U$, we approximate the effects of interactions using a self-consistent mean field, $U n_{i\uparrow} n_{i\downarrow}\approx U \langle n_{i\uparrow}\rangle n_{i\downarrow}+n_{i\uparrow}\langle n_{i\downarrow}\rangle +$ const.  As usual, we find the mean-field values $\langle n_{i\uparrow,\downarrow}\rangle$ by iteration until convergence, taking care to damp the update loop to avoid bistabilities in the solution. We perform the self-consistent calculation using a finite rescaling factor $\lambda$ for increased efficiency. We have checked that a rescaling $U'=U/\lambda$ results in $\lambda$-independent values of $U_c$ or spiral-F orders.

The calculation of the total electronic energy as a function of the polarization angle $\alpha_{M}$ between magnetic moments of adjacent AA regions, requires diagonalization of a supercell containing three minimal unit cells, the lattice vectors of the triangular superlattice are: 
\begin{equation}
\vec{T_1}=\vec{R_1}+\vec{R_2} = -r\vec{a_1}+(3 m_0+2r) \vec{a_2}
\end{equation}
\begin{equation}
\vec{T_2}= 2\vec{R_2}-\vec{R_1} = -(3m_o+2r)\vec{a_1}+(3 m_0+r) \vec{a_2}
\end{equation}
Since the diagonalization of the triangular superlattice is extremely time-consuming, our approach is to calculate self-consistently the magnetic moments contained within the minimal unit cell, and a  non-collinear mean-field Hamiltonian is constructed for the triple supercell, by rotation of the spins in the neighboring minimal cells by an angle $\alpha_{M}$:
\begin{equation}
H_{MF} = U\sum_{i,\sigma} \langle n_{i\sigma} \rangle n_{i\sigma^{\prime}} -\langle c^{\dagger}_{i\sigma} c_{i\sigma^{\prime}} \rangle c^{\dagger}_{i\sigma^{\prime}} c_{i\sigma} -E_{DC}
\end{equation}
where $E_{DC} = -U \left[ \langle n_{i\uparrow}\rangle \langle n_{i\downarrow}\rangle   -\langle c^{\dagger}_{i\downarrow} c_{i\uparrow} \rangle \langle c^{\dagger}_{i\uparrow} c_{i\downarrow} \rangle  \right] $  is a constant term, the new mean-values for the i-th site of this non-collinear Hamiltonian are calculated from the local magnetic moments from $
\langle n_{i\uparrow} \rangle=\frac{1}{2}(M^{i}_{0}+M^{i}_{z})$, $\langle n_{i\downarrow} \rangle=\frac{1}{2}(M^{i}_{0}-M^{i}_{z})$,$\langle c_{i\downarrow}^{\dagger} c_{i\uparrow}\rangle=\frac{1}{2}(M^{i}_{x}-iM^{i}_{y})$, $ \langle c_{i\uparrow}^{\dagger} c_{i\downarrow} \rangle=\frac{1}{2}(M^{i}_{x}+iM^{i}_{y})$. The minimal unit cell is chosen to be hexagonal and centered in the AA regions with vertices in the AB regions, since this geometry ensures that the rotations of the magnetic moments between adjacent minimall cells is carried out in an electronically depleted region, where the magnitude of the spins is negligible by comparison to the AA region . The non-collinear $H_{MF}$ is constructed for each value of $\alpha_{M}$ and the total electronic energy is calculated by direct diagonalization of the new Hamiltonian.

\bibliography{/Users/pablo/biblio}

\end{document}